\documentclass[5p,final,number,times]{elsarticle} 
\usepackage[titletoc,toc,title]{appendix}
\journal{Journal of Neuroscience Methods}
\renewcommand{\cite}[1]{\citep{#1}}
\usepackage{amsmath}

\usepackage{graphicx}
\usepackage{sidecap}

\usepackage[usenames,dvipsnames]{color}
\usepackage[colorlinks=true,urlcolor=Blue,citecolor=Violet,linkcolor=BrickRed]{hyperref}

\usepackage[displaymath,mathlines]{lineno}
\linenumbers
\nolinenumbers

\newcommand{\be}{\begin{linenomath}\begin{equation}}
    \newcommand{\ee}{\end{equation}\end{linenomath}}
\newcommand{\bea}{\begin{eqnarray}}
\newcommand{\eea}{\end{eqnarray}}

\newcommand{\units}[1]{{\,\rm #1}}

\newcommand{\fig}[1]{Fig.\,\ref{#1}}

\newcommand{\figs}[2]{Figs.\,\ref{#1}--\ref{#2}}
\newcommand{\figsand}[2]{Figs.\,\ref{#1} and \ref{#2}}

\newcommand{\eq}[1]{Eq.\,(\ref{#1})}
\newcommand{\Eq}[1]{Equation (\ref{#1})}

\def\d{{\rm d}}
\def\lD{\ell_D}
\def\lc{\ell_c}
\def\lg{\ell_g}
\def\tc{t_c}
\def\G{{\cal G}}
\def\D{{\cal D}}

\def\Tr{{\rm Tr\,}}
\def\qt{$q\mbox{--}t$ }

\begin{document}

\begin{frontmatter}

\title{Microstructure with Diffusion MRI: What Scale We Are Sensitive to?}

\author{Valerij G.\ Kiselev}
\cortext[mycorrespondingauthor]{Corresponding author}
\ead{kiselev@ukl.uni-freiburg.de}

\address[mymainaddress]{Medical Physics, Dpt. of Radiology, University Medical Center Freiburg, Faculty of Medicine, University of Freiburg, Freiburg, Germany}

\begin{abstract}
Diffusion-weighted MRI is the forerunner of the rapidly developed microstructural MRI aimed at in vivo evaluation of the cellular tissue architecture. This brief review focuses on the spatiotemporal scales of the microstructure that are accessible using different diffusion MRI techniques and the need to weight the measurability against the interpretability of results. Diffusion phenomena and models are first classified in two-dimensional space (the q-t-plane) of the measurement with narrow gradient pulses. Three-dimensional parameter space of the Stejskal--Tanner diffusion weighting adds more phenomena to this collection. Modern measurement techniques with larger number of parameters are briefly discussed under the overarching idea of diffusion weighting matching the geometry of the targeted cell species. 
\end{abstract}

\begin{keyword}
Diffusion; diffusion MRI; DWI; DTI; microstructure; cellular structure; microstructural MRI.
\end{keyword}

\end{frontmatter}

\section{Introduction}

The structure in the neural tissues spans a broad range of scales starting from about a micrometer (dendrites, axons) to some $20\units{\mu m}$ for large neurons. Intracellular structures are often finer while the cellular architecture can change on a larger scale (such as the cortical layers), which might be unresolvable in MR images. Evaluating deeply subresolution structures in vivo is an ambitious goal of modern MRI. Diffusion MRI is the forerunner of this development due to the fact that the MR-reporting molecules explore their local cellular environment with the typical size of $\lD\sim \sqrt{Dt}$, where $D$ is the diffusion coefficient and $t$ the diffusion time. Taking the typical value for water in the brain tissues, $D \sim 1\units{\mu m^2/ms}$ and the diffusion time from $1\units{ms}$ (ultimate oscillating-gradient measurements) to $1\units{s}$ (stimulated echo) gives the range from $1$ to about $30\units{\mu m}$, which is commensurate with the size of many cell species. 

\begin{figure}[tbp]
\includegraphics[width=\columnwidth]{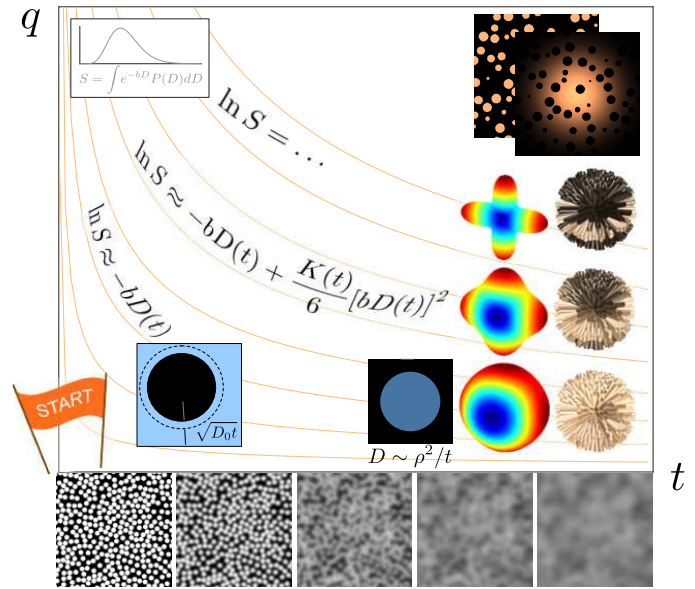}
\caption{An overview of diffusion measurements presented in the style of Ref.\cite{Novikov2018_models}. The schematics for the coarse-graining with increasing diffusion time is shown off axes, since this is the pure diffusion phenomenon, unrelated to MR measurements. See text for further explanations.  
\label{fig_map}}
\end{figure}

The above reason underscores the central role of scales for understanding what are the microstructural features a measurement is sensitive to. Discussing this question is the aim of the present paper. The first part of the discussion can be thought as an excursion in the landscape of diffusion weighted signal in the plane formed by the diffusion time $t$ and the wave vector $q$ of the gradient-induced magnetization \cite{Kiselev2017,Novikov2018_models}, \fig{fig_map}. Limiting the consideration to this plane is a large simplification, actually both quantities are estimates, respectively, of the duration and the magnitude of the continuous function 
\be \label{def_q} 
q(t) = \int_0^t \d t' g(t') \,, 
\ee
where $g(t)$ is the applied gradient of the Larmor frequency (the gradient of the main field multiplied with the gyromagnetic ratio of the MR-reporting spins). The value of $q$ to the end of diffusion weighting must be zero to avoid any interference between the diffusion weighting and the imaging. It is only the measurement with the ideally narrow gradient pulses for which a constant $q$ and the duration $t$ can be unambiguously defined. 

A well-known advantage of the narrow-pulse limit is that the signal is the Fourier transform, $G(t,q)$, of the sample-averaged diffusion propagator, $G(t,x)$ \cite{Callaghanbook}, 
\begin{linenomath}
\begin{align} 
S(t,q) 
&=\int \d^3 x_1 \frac{\d^3 x_0}{V} \, e^{iq(x_1-x_0)} \G(t,x_1,x_0) \nonumber \\
&= \int \d^3 x \, e^{iqx} G(t,x) \nonumber \\
&\equiv G(t,q)\,, \label{S=G}
\end{align}
\end{linenomath}
where $V$ is the sample volume and $\G(t,x_1,x_0)$ the exact diffusion propagator. In the spirit of this relation, a researcher measuring diffusion with a given $q$ anticipates the sensitivity to the spatial scale of the order of $1/q$. So, already in this simple example, the sensitivity to a given scale is understood in the spirit of the Fourier transform. 

Alike, this is applicable to the characteristic time a measurement is sensitive to. For measurements with arbitrary shaped weak gradients, the signal is dominated by the second-order term of the cumulant expansion \cite{Stepisnik93,Kiselev2010_diff_book}, 
\be \label{S=cum2}
\ln S 
\approx \int \d t_1 \d t_2 \, q(t_1)q(t_2) \D(t_2-t_1) 
= \int\frac{\d \omega}{2\pi}\, | q(\omega) |^2 \,\D(\omega) \,,
\ee
The quantity $\D(t)$ with its Fourier transform $\D(\omega)$ is the retarded (or causal) cumulant (autocorrelation function) of molecular velocity, $v(t)$ \cite{Stepisnik93,Novikov2010,Novikov2011_OG}:  
\be \label{def_vv}
\D(t) 
= \theta(t)\langle v(t) v(0) \rangle \,,
\ee
where $\theta(t)$ is the unit step function and the initial time moment $t=0$ is chosen arbitrary according to the time translation invariance, the physics synonym of the stationary process. Figure\,6 in \cite{Novikov2018_models} shows the relation of $\D$ to the conventional diffusion coefficient. For diffusion in media with no microstructure, such as homogeneous fluids, $\D(t)=D_0\delta(t)$ resulting in $\ln S \approx -bD_0$ with the standard definition of the b-value \cite{leBihan86}. \Eq{S=cum2} implies that  the sample-characterizing quantity $\D(\omega)$ defines the signal through the window of the measurement-representing function $| q(\omega) |^2$. While the time dependence $\D(t)$ is the major sample characteristic, it is sensed in the spirit of Fourier transform (cf.\ Refs.\,\cite{Stepisnik93,Novikov2011_OG,Kiselev2017,Novikov2018_models}).

\begin{figure}[tbp]
\includegraphics[width=\columnwidth]{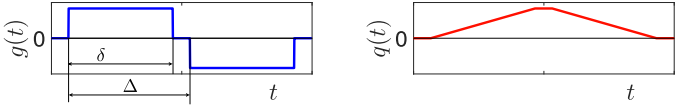}
\caption{The basic Stejskal--Tanner pulse sequence for the diffusion weighting \cite{Stejskal65}. The narrow pulse limit refers to $\delta\to 0$ and the opposite limit of wide pulses to $\delta = \Delta$. The refocusing pulse in the middle, which is applied in practical measurements, is taken into account by inverting the gradient preceding the refocusing pulse. 
\label{fig_grads}}
\end{figure}

Coming back to \eq{S=G}, the signal is maximum for the values of $q$ close to the inverse width of the diffusion propagator, which is of the order of $\lD\sim \sqrt{Dt}$. For very weak gradients, the exponential function can be approximated with unity and the signal is just the integral over the propagator, which is unity by the virtue of the particle conservation. The signal substantially depends on $q$ when $q\lD \sim 1$. By substituting $\lD$, this relation turns to $bD\sim 1$, which is the known rule for selecting the optimal b-value in the case of Gaussian diffusion. The scale $\lD$ can be thus called the default scale to which a measurement is sensitive. While this is optimal for measuring the value of $D$ in the case of Gaussian diffusion, the value of diffusivity, $D$, itself is rather unspecific to the microstructure if measured as it is commonly done for a given time moment \cite{Novikov2018_manifesto}. The following discussion is an attempt to answer the question about more microstructure-specific options and the signal penalty for deviating from the default values of $q$.

\section{Two-dimensional parameter space of narrow pulses}
\label{sec_NP}

\subsection{The cumulant expansion}
We now begin our excursion at the origin of the \qt plane, \fig{fig_map}. The signal at this point is maximum, $S=1$, which is the  normalization assumed hereafter. The origin is the highest point of the signal landscape. From that point, we see a downslope labelled with the contour lines of the b-value. For the narrow gradient pulses, $b=q^2t$, otherwise it is an order-of-magnitude estimate with the exact value defined by the specific gradient waveform. The b-value fully characterizes the signal attenuation in an aggregate of Gaussian non-exchangeable compartments. When applied to structured media such as biological tissues, the effect of diffusion weighting is irreducible to this single number being dependent on $q$ and $t$ separately. Therefore the actual contour lines of the signal do not coincide with the lines of constant $b$. In general, the signal is described by the cumulants of molecular displacements (or, equivalently of molecular velocity), with the leading term defined by the diffusion coefficient (or tensor in the general case), \eq{S=cum2}, the next term by the kurtosis etc., which are convolved with the increasing powers of $q(t)$ (or, equivalently, of $g(t)$) \cite{Jensen2005,Kiselev2010_diff_book}. This is often written in a simplified form using the proportionality $b\sim q^2$: 
\be \label{S=cums_b}
\ln S = -bD + C_2 b^2 + C_3 b^3 + \dots\,. 
\ee
This form however keeps hidden the fact that $D$ and all other coefficients depend on time, and, in more general terms, on the whole gradient waveform. Therefore, using this equation for analyzing the $b$-dependence is only reasonable when $b$ changes through the overall gradient magnitude while preserving the gradient waveform. With the increasing gradient, the contribution of the higher powers of $q(t)$ becomes dominant making the series divergent and thus the whole representation useless. 

\subsection{Measurability vs.\ interpretability}

It follows from \eq{S=cums_b} that accomplishing the apparently simple task of measuring the diffusion coefficient requires making a compromise between the measurability and the interpretability of the measurement. The optimal value of $b$ such as $bD\sim 1$ might be too large in media with non-Gaussian diffusion in which case the contribution of the higher-order terms in the series in \eq{S=cums_b} may result in a significant bias. Reducing $b$ helps, but decreases the effect of diffusion weighting, thus enhancing the noise propagation of the final result. 

Balancing the measurability vs.\ interpretability is a constant component of measurements. The penalty for such a compromise should be analysed in each specific case. Oftentimes, the shift toward larger b-values results in a smaller penalty than it might appear in view of the exponential signal decay in structureless fluids. In fact, the signal decrease for large $q$ in structured biological tissues is much slower, typically as an inverse power of $q$ as in examples considered below. 

\subsection{Time-dependent diffusivity}

We start descending from the highest point of the signal landscape in the direction of increasing time keeping the diffusion-weighing gradient small. In this regime, the signal is only sensitive to the diffusion tensor, which is represented in isotropic media by a single time-dependent quantity, the diffusion coefficient, $D(t)$. It is a decreasing function of time, since the microstructure in biological media restricts molecular diffusion, the longer time the more pronounced the effect is. In general, $D(t)$ levels off for long diffusion times in which case diffusion becomes effectively Gaussian by the virtue of the Central Limit Theorem (CLT). The caveat here is the need to realize what time can be considered sufficiently long. In a single compartment, it is $t\gg \tc$, where the correlation time, $\tc$, is the typical time of diffusion through a singe correlation length in the medium, $\lc$, for example defined by the size of cells. 

In a medium with multiple compartments, the CLT may apply to each compartment separately, but not yet to the whole medium if the exchange between the compartments is slow. This can be regarded as a constructive definition of compartments with account for the measurement technique. In many cases it coincides with the anatomical compartments, when exchange between them is slow \cite{Nilsson2013,Yang2018}. To give an example of the opposite, the water residence time inside the red blood cells is about $10\units{ms}$ \cite{Waldeck95}, which means that these cells can be treated as a compartment only when subjected to very fast measurements. The smallness of the exchange rate in the neural tissues enables the account for its effect using the simplified model of chemical exchange \cite{Kaerger85,Fieremans2010}. 


For short diffusion times, the initial decrease in $D(t)$ reveals the total surface area of impermeable interfaces in a unit volume of the medium, $A/V$, \cite{Mitra92} according to the relation 
\be \label{S=S/V}
D(t) = D_0 \left[ 1 - {\rm Const} {A\over V}\sqrt{D_0 t} + {\cal O} \left( \frac{D_0 t}{R_c^2}\right) \right] \,,
\ee
where $D_0$ is the bulk diffusivity, ${\rm Const} $ depends on the pulse sequence and the geometry of the interfaces \cite{Moutal2019_short_times} and the last term collects the corrections in the radius of curvature, $R_c$, and the surface relaxation \cite{Mitra93}. According to \eq{S=S/V}, the effect is the most pronounced when the diffusion length, $\lD=\sqrt{D_0t}$ is commensurate with the typical pore size, $V/A$. However, the tractability of the measurement requires selecting a shorter time, $\sqrt{D_0 t}\ll V/A$, to reduce the influence of the rest of the series in \eq{S=S/V}. In this regime, the measurement is practically insensitive to the curvature radius, $R_c\sim V/A$, due to the large difference in the scales, $\lD \ll R_c$. \Eq{S=S/V} has not found application to the brain due to the very small size of cells to which it could be applied. For example, application to white matter axons would require measurements with sub-millisecond durations, which are not feasible. 

In media with a microstructure, but unbounded space for diffusion, $D(t)$, reduces the most pronouncedly when the diffusion length is commensurate with the medium correlation length, $\lD\sim\lc$. The tractability of $D(t)$ in this regime is however low. In general, the medium structure gets increasingly blurred as seen by diffusing molecules, the process called the coarse-graining in physics, which is illustrated with a series of images of a granular two-dimensional medium in \fig{fig_map}. This process is discussed in more detail by \cite{Novikov2020_review}. In brief, for long diffusion times such that $\lD\gg\lc$, diffusion in a given compartment approaches its Gaussian asymptote as an inverse power of time, which reflects the effective dimensionality of diffusion and the statistics of long-distance correlations of the medium structure on the scale $\lD \gg \lc$ \cite{Novikov2014}. 


The above requirement of unbounded diffusion excluded the case of closed compartments, which is considered now. The property of being closed or open is irrelevant for short diffusion times for which \eq{S=S/V} equally applies. For long diffusion times, the size of the closed compartment, $\rho$, becomes relevant. Time is considered as long when $\rho \ll \lD$. According to this relation, the corresponding schematics in \fig{fig_map} is shown for rather long times although the real diffusion time might be as short as a couple of milliseconds in the case of very small compartments. For the narrow gradient pulses, \eq{S=G} predicts the signal in the form $S = 1 - {\cal O}(q^2 \rho^2)$. In more detail, it is $S = 1 - q^2 \langle \Delta x^2 \rangle /2$, where $\Delta x$ is the distance from the compartment's center of mass in the direction of $q$ and the averaging is made over the volume of the compartment. This relation follows from the exact diffusion propagator in the time-independent form $\G(t,x_1,x_0) = \eta(x_1)\eta(x_0)/V$, where $\eta(x)$ is the indicator function, which is unity inside the compartment and zero otherwise and $V$ the compartment volume. The diffusion coefficient found from this signal is $D = \langle \Delta x^2 \rangle /(2t)$, which coincides with the definition of $D$ via the mean squared displacement of diffusing molecules. 


The corner of long times and weak to moderate gradients in \fig{fig_map} is crowded because of the currently popular models of multiple Gaussian compartments \citep[section III]{Jelescu2017, Novikov2018_models}. Since the components of biological tissues are never really homogeneous, the assumption of Gaussian compartments relies on long diffusion times, in which case the CLT takes effect. This happens when $\lD\gg \lc$ within individual compartments. For such times, we can use the Gaussian approximation for small and moderate gradients. Deviation from this approximation for the whole multicompartment tissue is illustrated with the glyphs showing the increase in the angular resolution of crossing fibers with increasing $q$. As it follows from the cumulant expansion, the terms up to the order $q^n$ contain contributions of spherical harmonics up to the order $\ell = n$ \cite{Kiselev2010_diff_book,Novikov2018,Novikov2018_models}, whereas the Gaussian approximation is limited by $\ell =2$.  

Another set of images illustrates the model of white matter fibers as straight homogeneous cylinders \cite{Kroenke2004,Jespersen2007}, which are shown with a common origin to focus on their orientation distribution. This is in the core of what can be called the Standard Model of brain white matter \cite{Novikov2018_manifesto}. With the increasing $q$, less axons contribute to the powder-averaged signal, which results in the observed signal decrease in proportion to $1/\sqrt{b}$ \cite{Kroenke2004,McKinnon2017,Veraart2019,Veraart2020_axons}. For both examples, the probed orientation distribution does not refer to a specific length. However, diffusion length enters the result implicitly by defining the length of axons over which they should be effectively straight. The axon diameter falls out being too small for typical measurements in the brain requiring much larger gradients for been observable \cite{Novikov2014,Veraart2019}. Such gradients in combination with the larger axons in the spinal cord enabled ex vivo measurement of axonal diameter distribution \cite{Benjamini2016}. 

\subsection{Strong gradient}

The images in the upper right corner of the \qt plane illustrate the regime of long times and strong narrow gradients in which the signal is proportional to the medium correlation function. For diffusion in closed compartments, this phenomenon is commonly referred to as diffusion diffraction \cite{Callaghan91}.  The most pronounced effect is observed when $q\rho \sim 1$. 

The swiss-cheese-looking image in the same corner illustrates a similar phenomenon for diffusion in a connected porous medium \cite{Mitra92}. Such a medium is shaped by excluded volume of a complex geometry, typically MR-invisible (e.g., porous rock) filled with an MR-visible fluid (e.g., water) forming a single connected compartment with the complementary geometry. The key idea is that the diffusion propagator has the form of a nearly Gaussian envelope with a fine structure imprinted by the excluded, MR-invisible volume (this is the idea of the schematics in \fig{fig_map3d} showing the propagator magnitude). The propagator thus has two scales, $\lD=\sqrt{Dt}$ and $\lc$, respectively, with the relation $\lD \gg \lc$ as the condition for time to be treated as long. Substitution of this propagator in \eq{S=G} results in the signal equals to the convolution of a Gaussian propagator, $\exp(-Dtq^2$) with the medium correlation function, the latter arising from the indicator function, $\eta(x)$ of the pore space. In this way, we can access the scale of the medium correlation length, $\lc \ll \lD$. The price for that is the suppression of the informative part of the signal by the factor $(\lc/\lD)^d$ in $d$ spatial dimensions \cite{Mitra92}. 


We now return to the origin of the \qt plane to continue our excursion along the $q$ axis. Observing the decreasing signal we come to the domain of short times and strong gradients, which can be called the high-resolution limit \cite{Novikov2010}. The short times imply that the majority of spins explore only small volumes in which the diffusion coefficient is nearly constant. The signal is therefore simply $\langle \exp(-bD) \rangle$ where the averaging is performed over the distribution of $D$ in the sample \cite{Yablonskiy2003}. Note that the evaluation of the surface-to-volume ratio using low gradients \cite{Mitra92} can be understood as a particular case of such averaging. The signal in the high-resolution limit does not refer to a definite spatial scale because of the large difference, $\lD\ll \lc$. In media with multiple structural scales though, one has to realize that $D$ in the above expression is already averaged on all scales below $\lD$. 

\section{Three-dimensional parameter space of Stejskal--Tanner sequence}

\begin{figure}[tb]
\includegraphics[width=\columnwidth]{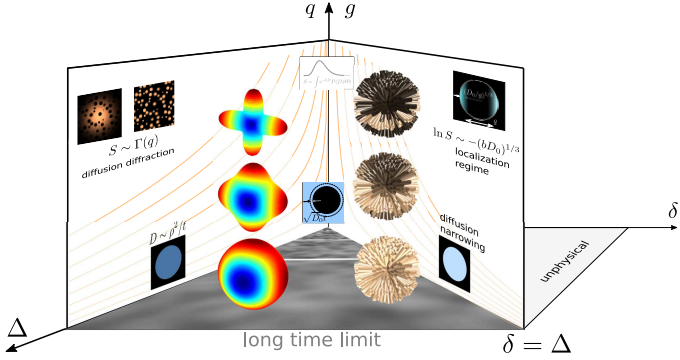}
\caption{Diffusion measurement regimes in three-dimensional parameter space of the Stejskal--Tanner pulse sequence. The left vertical plane corresponds to the narrow pulse limit with $\delta\to 0$, \fig{fig_map}. The right plane corresponds to the maximum pulse duration, $\delta=\Delta$, larger values of $\delta$ are impossible. The gradient strength in the vertical axis is labelled with both $q$ and $g$, since $q$ is the relevant quantity for $\delta\to 0$, while $g$ is constant, thus more convenient when $\delta=\Delta$. The contour lines of constant $b$ are different on these planes according to the exact expressions in terms of $q$ and $g$, respectively. The symbols shown in the planes stand for the phenomena that are specific to the corresponding limit of $\delta$. Those that are unspecific to the gradient waveform are shown in between the limiting planes. The coarse-graining symbolic from \fig{fig_map} is shown paving the floor, $q=g=0$, since it is the pure diffusion phenomenon, independent of the measurement technique. Further explanation in the text. 
\label{fig_map3d}}
\end{figure}

While the narrow pulse limit is lucid from the theory point of view, its practical realization is often difficult due to the hardware and biophysical limitations. Practically important are rather long gradient pulses, which are considered here in the limit of constantly applied gradients, $\delta=\Delta = t/2$ in the standard notation of the Stejskal -- Tanner measurement sequence \cite{Stejskal65}, \fig{fig_grads}. The parameter space of this sequence is three-dimensional comprising $q$, $\Delta$ and $\delta$ and we have to step away from the originally discussed \qt plane in the third orthogonal direction of $\delta$, \fig{fig_map3d}. 

When deployed in three dimensions, the measurement techniques shown in \fig{fig_map} behave differently. Those that are based on the Gaussian approximation to diffusion are largely insensitive to the gradient waveform because this dependence is absorbed in the numerical value of the b-factor. Such measurements are those for short diffusion times and based on the approximation of Gaussian compartments for long diffusion times, \fig{fig_map3d}. In contrast, the measurements inspired by \eq{S=G} require narrow gradient pulses such that $\delta \ll \tc$. The reason is the approximate solution to the Bloch--Torrey equation that describes the evolution of the transverse magnetization due to diffusion and precession. In other words it describes the complete interplay between diffusion of molecules and their nuclear magnetization \cite{Torrey56}. The narrow pulses enable approximating the gradient action with the instant phase acquisition, thus decoupling the gradient effect from molecular diffusion.

Two measurements are shown in \fig{fig_map3d} that are specific to the limit of long gradient pulses, $\delta = \Delta$. Schematics of long-time diffusion in a closed compartment, $\lD \gg \rho$ is shown in the corner of long times and weak gradients. The effect of the diffusion weighting is much weaker than for the narrow pulses, 
\be \label{Wayne}
S - 1 
\sim \frac{\rho^4g^2t}{D_0}
\sim  q^2 \rho^2 \frac{\rho^2}{D_0 t} \,,
\ee
where $q$ is estimated as $q\sim gt$, while $q^2\rho^2$ is the estimate for the effect of narrow pulses, and the last factor $\rho^2/(D_0t) = (\rho/\lD)^2\ll 1$. The same small parameter spoils calculation of diffusion coefficient using the standard procedure of dividing the signal, \eq{Wayne} by the b-factor, $b\sim g^2t^3 \sim q^2t$. This results in $D_{\rm app} \sim (\rho^2/t)(\rho^2 /D_0 t)$, where the first factor is the true diffusion coefficient in a small compartment. The reason for this effect is the diffusion narrowing: Due to the relatively fast diffusion, the distribution of spin phases gets narrower thus reducing the signal attenuation \cite{Robertson66, Wayne66, Neuman74, vanGelderen94_diff}. Diffusion narrowing has been actively studied in the context of transverse relaxation, see  \cite{Kiselev2018_review} for a review.

The comparison with narrow pulses is opposite in the same system measured with strong constantly applied gradients, the so-called localization regime \cite{Stoller91, Huerlimann95, Moutal2019_localization}. While this phenomenon is illustrated in \fig{fig_map3d} with the signal from a spherical cavity, its essence was derived by \cite{Stoller91} for a flat impermeable interface orthogonal to the applied gradient. The major signal comes from the layer adjacent to the interfaces with the layer thickness about $\lg = (D/g)^{1/3}$. Since the flat interface does not introduce any length scale, there is no medium-related size entering the dominant signal contribution. The strong gradient is defined by the condition $\lg \ll \lD$, which is equivalent to $bD \gg 1$ with $bD \sim Dg^2t^3$. In contrast to the acquition with the narrow pulses, the signal is exponentially suppressed although less in comparison with the free diffusion, $\ln S \sim - (bD)^{1/3}$. Medium-related length enters the pre-exponential factor for media with curved interfaces such as shown in \fig{fig_map3d} reflecting the fraction of spins affected by the suitably oriented interfaces \cite{Huerlimann95, Moutal2019_localization}.

\section{Advanced measurement techniques: Matching the cell geometry }

Diffusion weighting techniques that are in the today's research focus have more controlling parameters than the Steiskal--Tanner waveform, which makes unfeasible any graphical classification similar to \fig{fig_map3d}. In particular, there are numerous methods of combining gradient pairs, which can be loosely considered as successors of the double-diffusion encoding \cite{Mitra95, Shemesh2016}. Combining gradient pairs can be extended to the continuum limit of constantly changing magnitude and direction of the gradient vector, $g(t)$, \fig{fig_gradnd} \cite{Eriksson2013, Lasic2014, Szczepankiewicz2015}). The unifying idea of those methods can be called the ``geometry matching" that is relying to the distinct shapes of tissue compartments to get more specific information about them. This is a relatively new trend in microstructural MRI as compared with the traditional approach of obtaining the tissue-averaged signal characteristics. 

\subsection{Weak gradients: Why diffusion tensor is not a model}

\begin{figure}[tbp]
\includegraphics[width=\columnwidth]{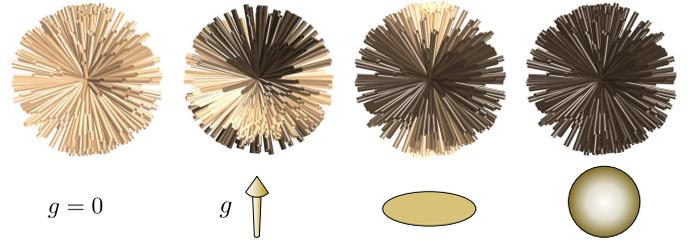}
\caption{Basic forms of diffusion weighting (the gradient schematics in the bottom row) illustrated with the effect on the isotropic distribution of axons that are shown with the common origin for clarity, dark color for suppressed signal. From left to right: No weighting, linear, planar, and spherical encodings. 
\label{fig_gradnd}}
\end{figure}

As the rule, geometry matching requires strong diffusion weighting, with the obvious penalty of a significant signal loss. The reason is that the signal for weak gradients is not very informative. It is dominated by the tissue-averaged diffusion tensor, $D_{ij}$, where the indices $i,j = 1,2,3$, and the most relevant outcome of the complex gradient shape is the b-matrix, $b_{ij}$. The signal is then 
\be\label{S=bD}
\ln S \approx -b_{ij}D_{ij} = -\Tr bD\,,
\ee 
where the summation over the repeating indices is implied. Since this is linear in $b$, the effect of any b-matrix can be represented as the sum of the following three ones that correspond to the three images for non-zero gradients in \fig{fig_gradnd}. In the basis of their eigenvectors
\be\label{bdiag}
b_{\rm 1d} = \begin{pmatrix}
0 & 0 & 0\\
0 & 0 & 0\\
0 & 0 & 1
\end{pmatrix} ,\,
b_{\rm 2d} = \frac{b}{2}\begin{pmatrix}
0 & 0 & 0\\
0 & 1 & 0\\
0 & 0 & 1
\end{pmatrix} ,\,
b_{\rm 3d} = \frac{b}{3}\begin{pmatrix}
1 & 0 & 0\\
0 & 1 & 0\\
0 & 0 & 1
\end{pmatrix} .
\ee
These matrices realize the linear, planar and spherical diffusion encodings, respectively. Obviously, the first two select a preferred direction in space, the third does not.

According to \eq{S=bD}, the usage of linear and spherical encodings with weak gradients does not add any new information. It solely changes the combination of the diffusion tensor components that determines the measured signal. Looking broadly, while diffusion tensor for a given time moment can indicate the overall (macroscopic) tissue anisotropy, it is not specific to the microstructure. Even such a pronounced structure as a fiber crossing is not resolvable with only $D_{ij}$ as illustrated with the low-q glyph in \figsand{fig_map}{fig_map3d}. The time dependence, $D(t)$ is more informative, as is the dependence on the angle between the two gradient pairs of the double diffusion encoding for short mixing times \cite{Mitra95,Shemesh2016}, \fig{fig_dPFG}. However, it turned out that these apparently different dependencies contain the same information about the microstructure \cite{Jespersen2012_dPFG}. 

Making a short break for discussing the terminology, the diffusion and kurtosis tensors were classified as signal representations as contrasted with parameters of microstructural models \cite{Novikov2018_manifesto}. In brief, the reason was the universal signal dependence on $b$ in terms of  these parameters, cf.\,\eq{S=cums_b}, and the low specificity of such representation to the microstructure. As a rule of thumb, representations are mathematical expressions for describing the signal, while models include pictures of the most relevant tissue properties, \figs{fig_gradnd}{fig_FEXI} can serve as simple examples.

\subsection{Strong gradients: Accessing the microstructure}

\begin{figure}[tbp]
\includegraphics[width=\columnwidth]{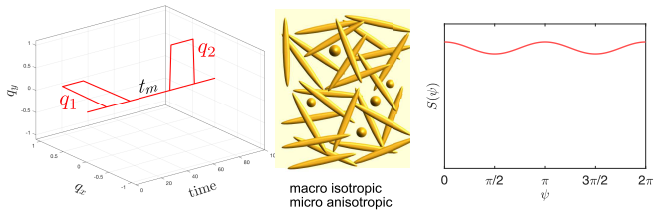}
\caption{Schematic of the double diffusion encoding. Diffusion weighting consists of two gradient pairs that induce two successively applied wave vectors $q_1$ and $q_2$ with the separation $t_m$ and a variable angle between them ($\psi=\pi/2$ in the left image). The signal obtained with such a sequence depends on $\psi$ when acquired in a medium built with anisotropic cells (middle image). The anticipated signal variation in the limit of long $t_m$ is proportional to $\cos 2\psi$ (right panel). Qualitatively similar effect for short $t_m$, for which the signal is proportional to $\cos \psi$, does not contain any new information as compared with the time-dependent diffusion coefficient \cite{Jespersen2012_dPFG}. \label{fig_dPFG}}
\end{figure}

The first example considered here is the detection of anisotropic microscopic compartments such as cell processes, in tissues that can be apparently isotropic due to the random orientations of such compartments. This anisotropy, often hidden on the macroscopic level of imaging is referred to as the microscopic anisotropy (alternative terminology is discussed in \cite{Shemesh2016}). Historically the first approach was the double diffusion encoding \cite{Mitra95}, which has been applied to ex-vivo specimens \cite{Komlosh2007}, live animal \cite{Shemesh2012} and human brain  \cite{Lawrenz2013, Lawrenz2019} naming only the first observations (\fig{fig_dPFG}). Proper characterization of the microscopic compartments refers the their rotationally-invariant parameters, which are independent of the measurement setup and the sample orientation \cite{Jespersen2013}. 

\begin{figure}[tbp]
\includegraphics[width=\columnwidth]{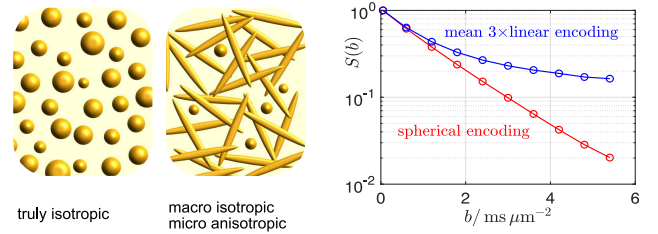}
\caption{Spherical encoding can be used to detect the micro-anisotropy in macroscopically isotropic samples. For a truly isotropic medium (the left image), there is no qualitative difference between the signals obtained with the both methods. The linearly encoded signal from microscopically anisotropic medium (the middle image) levels off for large $b$ because of subpopulation of cells that are nearly orthogonal to the gradient thus escaping the signal suppression (blue line). This is not the case for the spherical encoding (red line). The data were obtained by averaging the signal from 707 white matter voxels in a normal human brain \cite{Dhital2018_iso}. The small deviation from the exponential decay for large $b$ limits the presence of small round cells, which otherwise would result in the signal levelling off similar to the blue line. Note the coincidence of both signals for low $b$ in agreement with the discussion after \eq{bdiag}. 
\label{fig_iso}}
\end{figure}

Microscopic anisotropy can be detected by comparison of the traditional linear diffusion encoding with the spherical encoding (or, synonymously, isotropic weighting) with $b_{\rm 3d}\sim {\rm diag\,}[1,1,1]$, \eq{bdiag}, which is illustrated in \fig{fig_iso} \cite{Lasic2014, Szczepankiewicz2015, Lampinen2017}. The major difference between the linear and spherical encodings is that the linear encoding suppresses the signal from molecules that can move in the direction of the applied gradient, while the spherical encoding suppresses the signal from molecules that are mobile in at least one direction. While the principle of spherical encoding is clear, quantitation of results obtained in media with microstructure requires caution.  The accuracy with which the diffusion weighting is isotropic depends on both the pulse sequence and the sample according to the three-dimensional version of \eq{S=cum2} \cite{Jespersen2014, Jespersen2019, Moutal2019_short_times}.

Spherical encoding was also used to suppress the signal from all tissue compartments except for small compact cells (the so-called dot compartment) with the aim to evaluate their weight in the total signal \cite{Dhital2018_iso,Tax2020_dot} (\fig{fig_iso}). The result showed very little amount of such cells in the brain except some small, but observable fraction in the cerebellum \cite{Tax2020_dot}.

\begin{figure}[tbp]
\includegraphics[width=\columnwidth]{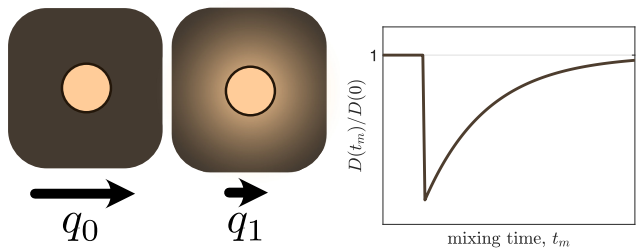}
\caption{Schematic of the exchange measurement using a mobility filter. Left: Water inside a small cell escapes the signal suppression with a pair of strong gradient characterised by the wave vector $q_0$. Diffusion coefficient is measured with a small wave vector $q_1$. The value of $D$ drops right after the suppression to the value inside the small cell. Due to the exchange, molecules with unsuppressed signal move to the interstitium where they have a larger diffusivity. This results in a gradual recovery of the initial diffusion coefficient as a function of the so-called mixing time, $t_m$, after the mobility filter application. 
\label{fig_FEXI}}
\end{figure}

The little signal suppression in the  compartments with small size in the direction of a strong applied gradient was used for evaluating the water exchange rate between such compartments and the interstitium \cite{Aslund2009, Lasic2011_exchange, Lampinen2016, Lasic2016}. The principle of this measurement is illustrated in \fig{fig_FEXI}. The resulted residence time of a few hundreds of milliseconds agrees with measurements in the culture of neural cells \cite{Yang2018}. 

The whole-brain averaged intra-axonal diffusivity was measured in vivo using the planar weighting with $b\sim {\rm diag\,}[1,1,0]$ for effective narrowing of the native orientation distribution of white matter axons and suppressing the signal from other cells (the third panel in \fig{fig_gradnd}). The aim was to measure the averaged intra-axonal diffusivity in living human brain with the resulted value of $2.25\pm 0.03\units{\mu m^2/ms}$ \cite{Dhital2019}. 

For the powder-averaged signal, the planar encoding is complementary to the more traditional weighting with a linear gradient (\fig{fig_gradnd}). With this technique, the axonal orientation distribution is effectively isotropic and a strong gradient suppresses the signal from all axons, but those that are nearly orthogonal to the gradient direction. The fraction of such axons decreases in proportion to $1/\sqrt{bD}$, which defines the signal dependence on $b$ \cite{Jensen2016, McKinnon2017, Veraart2019, Veraart2020_axons}. Marginal deviations from this dependence signify the limits of modeling the axons as straight infinitely thin cylinders \cite{Veraart2020_axons}. This approach is discussed in more detail by \cite{Novikov2020_review}.

\section{Interplay with microscopic magnetic field}

Quantitative diffusion measurements face several experimental challenges related to the monitoring the real gradient strength and time course. A similar problem appears on the level of microstructure in the presence of heterogeneous magnetic susceptibility inside the sample \cite{Huerlimann98}, in the biomedical context, due to, e.g., deoxygenated blood, contrast agent or iron deposition in a specific cell population. The resulted microscopic magnetic field accelerates the transverse relaxation (see \cite{Kiselev2019} and \cite{Kiselev2018_review} for recent reviews on these subjects, respectively). The same microscopic magnetic field interplays with the applied diffusion-weighting gradients thus affecting the measurement results. The effect depends on the scale of the microscopic field $\ell_B$, relative to the diffusion length, $\lD$. Since the elementary magnetic dipole field is scaleless, the former characteristic length is defined by the correlation length of the field-inducing structure (this property was called locality \cite{Kiselev2002, Novikov2008}). 

In the case of relatively large magnetic structure, $\ell_B \gg \lD$, each spin moves in a constant magnetic field gradient although different for different regions. The diffusion weighting to the first, Gaussian approximation is simply $S = \langle e^{-bD} \rangle$, where $b$ has a random component due to the additional local gradient and the averaging is performed over its spatial distribution. Since the exponential function is convex, this increases the signal as compared with the expected value, $S > e^{-b_0 D}$, where $b_0$ is the intended b-factor. In other words, areas where the macroscopic gradient partially compensates the applied one increase the signal more than the decrease due to areas where the microscopic gradient with the same magnitude adds to the applied one. The genuine diffusion coefficient is thus underestimated \cite{Does99, Zhong91, Kiselev2004}. 

The picture is less transparent for the case of equal scales,  $\ell_B \sim \lD$. Truly bipolar gradient weighting (no refocusing pulse) changes the effect sign. In contrast, diffusion weighting with the refocusing results in the underestimation of diffusion coefficient with the magnitude dependent on the gradient time course and statistics of the spatial organization of microscopic magnetic susceptibility. Being not overall large, this effect should be taken into account when dealing with subtle phenomena in diffusion measurements \cite{Novikov2018_ackerman}.

\section{Concluding remarks: Microstructural MRI as a physics discipline}

Physics can be regarded as a science about spatio-temporal scales and transitions between them. Making another excursion, now through the scales, let's begin at the fundamental level of elementary particles. Elementary particle physics has accumulated high-precision knowledge about the particle properties on the subnuclear scales. The challenge of calculating the mass and the gyromagnetic ratio of the proton is however still challenging the researchers. Moving one level up, these difficulties pose no problem for atomic and molecular physics for which the proton is a point-like particle with the empirically defined mass, charge, spin and magnetic moment. Molecular physics studies extremely complex phenomena on the scale around a nanometer, in particular the molecular motion in the liquid state. However, the challenges of this discipline do not affect the MRI community. The information we need from the molecular scale is the values of the relaxation rates, $R_1$ and $R_2$ and the bulk diffusion coefficient. 

These examples illustrate one of the fundamental notion of renormalizability, which in the broad sense refers to the possibility to include all effects of finer spatial scales in a handful of effective measurable parameters when considering physics on coarser scales. This is why atomic physics does not depend on the resolution of the long-standing problems of elementary particle physics and, in turn, diffusion MRI does not depend on the attempts to calculate the molecular diffusion coefficient from the first principle.  

While the above examples are obvious, it is not always clear what are the effective parameters for the considered phenomenon. This problem was especially challenging for physics of critical phenomena when quite different physical systems demonstrate largely universal behavior of their thermodynamic properties. One of the main achievements in physics in the second half of the 20th century was the development of methods for finding such parameters, called relevant, that are present after the transition from the molecular to the macroscopic scale. 

Somewhat toned down, this problem reincarnates in microstructural MRI as the question which cellular features are still measurable after the coarse graining to the MRI voxels. The main conclusion from the above discussion is that there is no universal answer to this question, For example, the surface to volume ratio might be the relevant parameter for short diffusion time while the long-distance structural correlations and the effective dimensionality of diffusion compartments are relevant for long times. The range of spatial scales covered by diffusion MRI is about the factor of 30 as indicated by the available values of the diffusion length, $\lD$. This is much smaller than for example the gap of five orders of magnitude between the elementary particle and atomic physics. In MRI, we typically do not work so deeply in the limiting cases, which affects the accuracy of our approximations. This accuracy depends on the relation between the measurement-defined scale, $\lD$, and the structural scale of the measured sample, $\lc$. Due to the large variability in the latter, the ratio might change in a larger range than $\lD$, which effectively enlarge the scope of accessible scales. In addition to the diffusion time, the gradient strength and waveform can radically change the relevant parameters thus further enlarging the palette of possibilities in the exciting research field of microstructural MRI. 

\section*{Acknowledgement} 
Im grateful to Dmitry Novikov for stimulating discussions. 

Declarations of competing interest: none.

\bibliographystyle{elsarticle-num} 
\bibliography{literatureMRI}

\begin{thebibliography}{10}
\expandafter\ifx\csname url\endcsname\relax
  \def\url#1{\texttt{#1}}\fi
\expandafter\ifx\csname urlprefix\endcsname\relax\def\urlprefix{URL }\fi
\expandafter\ifx\csname href\endcsname\relax
  \def\href#1#2{#2} \def\path#1{#1}\fi

\bibitem{Kiselev2017}
V.~G. Kiselev, Fundamentals of diffusion {MRI} physics, NMR Biomed 30~(3)
  (2017) e3602.
\newblock \href {http://dx.doi.org/10.1002/nbm.3602}
  {\path{doi:10.1002/nbm.3602}}.

\bibitem{Novikov2018_models}
D.~S. Novikov, E.~Fieremans, S.~N. Jespersen, V.~G. Kiselev, Quantifying brain
  microstructure with diffusion {MRI: T}heory and parameter estimation, NMR
  Biomed 32~(4) (2018) e3998.
\newblock \href {http://dx.doi.org/10.1002/nbm.3998}
  {\path{doi:10.1002/nbm.3998}}.

\bibitem{Callaghanbook}
P.~T. Callaghan, Principles of nuclear magnetic resonance microscopy, Oxford
  University Press Inc., New York, 1991.

\bibitem{Stepisnik93}
J.~Stepi\u{s}nik, Time-dependent self-diffusion by {NMR} spin-echo, Physica B
  183 (1993) 343--350.

\bibitem{Kiselev2010_diff_book}
V.~G. Kiselev, The cumulant expansion: An overarching mathematical framework
  for understanding diffusion {NMR}, in: D.~K. Jones (Ed.), Diffusion {MRI}:
  Theory, Methods and Applications, Oxford University Press, New York, 2010,
  Ch.~10, pp. 152--168.

\bibitem{Novikov2010}
D.~S. Novikov, V.~G. Kiselev, Effective medium theory of a diffusion-weighted
  signal, NMR Biomed 23~(7) (2010) 682--97.
\newblock \href {http://dx.doi.org/10.1002/nbm.1584}
  {\path{doi:10.1002/nbm.1584}}.

\bibitem{Novikov2011_OG}
D.~S. Novikov, V.~G. Kiselev, Surface-to-volume ratio with oscillating
  gradients, J Magn Reson 210~(1) (2011) 141--145.
\newblock \href {http://dx.doi.org/10.1016/j.jmr.2011.02.011}
  {\path{doi:10.1016/j.jmr.2011.02.011}}.

\bibitem{leBihan86}
D.~Le~Bihan, E.~Breton, D.~Lallemand, P.~Grenier, E.~Cabanis, M.~Laval-Jeantet,
  Mr imaging of intravoxel incoherent motions: application to diffusion and
  perfusion in neurologic disorders, Radiology 161~(2) (1986) 401--7.
\newblock \href {http://dx.doi.org/10.1148/radiology.161.2.3763909}
  {\path{doi:10.1148/radiology.161.2.3763909}}.

\bibitem{Stejskal65}
E.~O. Stejskal, J.~E. Tanner, Spin diffusion measurements: {S}pin echoes in the
  presence of a time‐dependent field gradient, The Journal of Chemical
  Physics 42~(1) (1965) 288--292.
\newblock \href {http://arxiv.org/abs/http://dx.doi.org/10.1063/1.1695690}
  {\path{arXiv:http://dx.doi.org/10.1063/1.1695690}}, \href
  {http://dx.doi.org/10.1063/1.1695690} {\path{doi:10.1063/1.1695690}}.

\bibitem{Novikov2018_manifesto}
D.~S. Novikov, V.~G. Kiselev, S.~N. Jespersen, On modeling, Magn Reson Med
  79~(6) (2018) 3172--3193.
\newblock \href {http://dx.doi.org/10.1002/mrm.27101}
  {\path{doi:10.1002/mrm.27101}}.

\bibitem{Jensen2005}
J.~H. Jensen, J.~A. Helpern, A.~Ramani, H.~Lu, K.~Kaczynski, Diffusional
  kurtosis imaging: The quantification of non-gaussian water diffusion by means
  of magnetic resonance imaging, Magn. Reson. Med. 53 (2005) 1432--1440.

\bibitem{Nilsson2013}
M.~Nilsson, J.~L{\"a}tt, D.~van Westen, S.~Brockstedt, S.~Lasi\v{c},
  F.~St{\aa}hlberg, D.~Topgaard, Noninvasive mapping of water diffusional
  exchange in the human brain using filter-exchange imaging, Magn Reson Med
  69~(6) (2013) 1573--81.
\newblock \href {http://dx.doi.org/10.1002/mrm.24395}
  {\path{doi:10.1002/mrm.24395}}.

\bibitem{Yang2018}
D.~M. Yang, J.~E. Huettner, G.~L. Bretthorst, J.~J. Neil, J.~R. Garbow,
  J.~J.~H. Ackerman, Intracellular water preexchange lifetime in neurons and
  astrocytes, Magn Reson Med 79~(3) (2018) 1616--1627.
\newblock \href {http://dx.doi.org/10.1002/mrm.26781}
  {\path{doi:10.1002/mrm.26781}}.

\bibitem{Waldeck95}
A.~R. Waldeck, M.~H. Nouri-Sorkhabi, D.~R. Sullivan, P.~W. Kuchel, Effects of
  cholesterol on transmembrane water diffusion in human erythrocytes measured
  using pulsed field gradient {NMR}, Biophys Chem 55~(3) (1995) 197--208.

\bibitem{Kaerger85}
J.~K\"arger, {NMR} self-diffusion studies in heterogeneous systems, Adv.
  Colloid and Interface Sci. 23 (1985) 129--148.

\bibitem{Fieremans2010}
E.~Fieremans, D.~S. Novikov, J.~H. Jensen, J.~A. Helpern, Monte carlo study of
  a two-compartment exchange model of diffusion, NMR Biomed 23~(7) (2010)
  711--24.
\newblock \href {http://dx.doi.org/10.1002/nbm.1577}
  {\path{doi:10.1002/nbm.1577}}.

\bibitem{Mitra92}
P.~Mitra, P.~Sen, L.~Schwartz, P.~Le~Doussal, Diffusion propagator as a probe
  of the structure of porous media, Phys Rev Lett 68~(24) (1992) 3555--3558.
\newblock \href {http://dx.doi.org/10.1103/PhysRevLett.68.3555}
  {\path{doi:10.1103/PhysRevLett.68.3555}}.

\bibitem{Moutal2019_short_times}
N.~Moutal, I.~Maximov, D.~Grebenkov, Probing surface-to-volume ratio of an
  anisotropic medium by diffusion {NMR} with general gradient encoding, IEEE
  Transactions on Medical Imaging (2019) 1--1\href
  {http://dx.doi.org/10.1109/TMI.2019.2902957}
  {\path{doi:10.1109/TMI.2019.2902957}}.

\bibitem{Mitra93}
P.~Mitra, P.~Sen, L.~Schwartz, Short-time behavior of the diffusion coefficient
  as a geometrical probe of porous media, Phys Rev B Condens Matter 47~(14)
  (1993) 8565--8574.
\newblock \href {http://dx.doi.org/10.1103/physrevb.47.8565}
  {\path{doi:10.1103/physrevb.47.8565}}.

\bibitem{Novikov2020_review}
D.~S. Novikov, J Neurosci Methods this issue.

\bibitem{Novikov2014}
D.~S. Novikov, J.~H. Jensen, J.~A. Helpern, E.~Fieremans, Revealing mesoscopic
  structural universality with diffusion, Proc Natl Acad Sci U S A 111~(14)
  (2014) 5088--93.
\newblock \href {http://dx.doi.org/10.1073/pnas.1316944111}
  {\path{doi:10.1073/pnas.1316944111}}.

\bibitem{Jelescu2017}
I.~O. Jelescu, M.~D. Budde, Design and validation of diffusion {MRI} models of
  white matter, Front Phys 28.
\newblock \href {http://dx.doi.org/10.3389/fphy.2017.00061}
  {\path{doi:10.3389/fphy.2017.00061}}.

\bibitem{Novikov2018}
D.~S. Novikov, J.~Veraart, I.~O. Jelescu, E.~Fieremans, Rotationally-invariant
  mapping of scalar and orientational metrics of neuronal microstructure with
  diffusion {MRI}, Neuroimage 174 (2018) 518--538.
\newblock \href {http://dx.doi.org/10.1016/j.neuroimage.2018.03.006}
  {\path{doi:10.1016/j.neuroimage.2018.03.006}}.

\bibitem{Kroenke2004}
C.~D. Kroenke, J.~J.~H. Ackerman, D.~A. Yablonskiy, On the nature of the {NAA}
  diffusion attenuated {MR} signal in the central nervous system, Magn Reson
  Med 52 (2004) 1052--1059.

\bibitem{Jespersen2007}
S.~N. Jespersen, C.~D. Kroenke, L.~{\O}stergaard, J.~J.~H. Ackerman, D.~A.
  Yablonskiy, Modeling dendrite density from magnetic resonance diffusion
  measurements, NeuroImage 34~(4) (2007) 1473--86.
\newblock \href {http://dx.doi.org/10.1016/j.neuroimage.2006.10.037}
  {\path{doi:10.1016/j.neuroimage.2006.10.037}}.

\bibitem{McKinnon2017}
E.~T. McKinnon, J.~H. Jensen, G.~R. Glenn, J.~A. Helpern, Dependence on b-value
  of the direction-averaged diffusion-weighted imaging signal in brain, Magn
  Reson Imaging 36 (2017) 121--127.
\newblock \href {http://dx.doi.org/10.1016/j.mri.2016.10.026}
  {\path{doi:10.1016/j.mri.2016.10.026}}.

\bibitem{Veraart2019}
J.~Veraart, E.~Fieremans, D.~S. Novikov, On the scaling behavior of water
  diffusion in human brain white matter, NeuroImage 185 (2019) 379--387.
\newblock \href {http://dx.doi.org/10.1016/j.neuroimage.2018.09.075}
  {\path{doi:10.1016/j.neuroimage.2018.09.075}}.

\bibitem{Veraart2020_axons}
J.~Veraart, D.~Nunes, U.~Rudrapatna, E.~Fieremans, D.~K. Jones, D.~S. Novikov,
  N.~Shemesh, Noninvasive quantification of axon radii using diffusion mri,
  eLife 9 (2020) e49855.
\newblock \href {http://dx.doi.org/10.7554/eLife.49855}
  {\path{doi:10.7554/eLife.49855}}.

\bibitem{Benjamini2016}
D.~Benjamini, M.~E. Komlosh, L.~A. Holtzclaw, U.~Nevo, P.~J. Basser, White
  matter microstructure from nonparametric axon diameter distribution mapping,
  NeuroImage 135 (2016) 333--344.
\newblock \href {http://dx.doi.org/10.1016/j.neuroimage.2016.04.052}
  {\path{doi:10.1016/j.neuroimage.2016.04.052}}.

\bibitem{Callaghan91}
P.~T. Callaghan, A.~Coy, D.~Macgowan, K.~J. Packer, F.~O. Zelaya,
  Diffraction-like effects in {NMR} diffusion studies of fluids in porous
  solids, Nature 351~(6326) (1991) 467--469.

\bibitem{Yablonskiy2003}
D.~A. Yablonskiy, G.~L. Bretthorst, J.~J. Ackerman, Statistical model for
  diffusion attenuated mr signal, Magn Reson Med 50~(4) (2003) 664--669.
\newblock \href {http://dx.doi.org/10.1002/mrm.10578}
  {\path{doi:10.1002/mrm.10578}}.

\bibitem{Torrey56}
H.~C. Torrey, Bloch equations with diffusion terms, Phys. Rev. 104 (1956) 563.

\bibitem{Robertson66}
B.~Robertson, Spin-echo decay of spins diffusing in a bounded region, Phys.
  Rev. 151 (1966) 273--277.
\newblock \href {http://dx.doi.org/10.1103/PhysRev.151.273}
  {\path{doi:10.1103/PhysRev.151.273}}.

\bibitem{Wayne66}
R.~C. Wayne, R.~M. Cotts, Nuclear-magnetic-resonance study of self-diffusion in
  a bounded medium, Phys. Rev. 151 (1966) 264--272.
\newblock \href {http://dx.doi.org/10.1103/PhysRev.151.264}
  {\path{doi:10.1103/PhysRev.151.264}}.

\bibitem{Neuman74}
C.~H. Neuman, Spin echo of spins diffusing in a bounded medium, Journal of
  Chemical Physics 60~(11) (1974) 4508--4511.
\newblock \href {http://dx.doi.org/10.1063/1.1680931}
  {\path{doi:10.1063/1.1680931}}.

\bibitem{vanGelderen94_diff}
P.~van Gelderen, D.~DesPres, P.~C. van Zijl, C.~T. Moonen, Evaluation of
  restricted diffusion in cylinders. phosphocreatine in rabbit leg muscle, J
  Magn Reson B 103~(3) (1994) 255--60.
\newblock \href {http://dx.doi.org/10.1006/jmrb.1994.1038}
  {\path{doi:10.1006/jmrb.1994.1038}}.

\bibitem{Kiselev2018_review}
V.~G. Kiselev, D.~S. Novikov, Transverse {NMR} relaxation in biological
  tissues, NeuroImage 182 (2018) 149--168.
\newblock \href {http://dx.doi.org/10.1016/j.neuroimage.2018.06.002}
  {\path{doi:10.1016/j.neuroimage.2018.06.002}}.

\bibitem{Stoller91}
F.~Stoller, W.~Happer, F.~Dyson, Transverse spin relaxation in inhomogeneous
  magnetic fields, Phys Rev A 44~(11) (1991) 7459--7477.

\bibitem{Huerlimann95}
M.~Hurlimann, K.~Helmer, T.~Deswiet, P.~Sen, Spin echoes in a constant gradient
  and in the presence of simple restriction, Journal of Magnetic Resonance,
  Series A 113~(2) (1995) 260 -- 264.
\newblock \href {http://dx.doi.org/https://doi.org/10.1006/jmra.1995.1091}
  {\path{doi:https://doi.org/10.1006/jmra.1995.1091}}.

\bibitem{Moutal2019_localization}
N.~Moutal, K.~Demberg, D.~S. Grebenkov, T.~A. Kuder, Localization regime in
  diffusion {NMR: T}heory and experiments, J Magn Reson 305 (2019) 162--174.
\newblock \href {http://dx.doi.org/10.1016/j.jmr.2019.06.016}
  {\path{doi:10.1016/j.jmr.2019.06.016}}.

\bibitem{Mitra95}
P.~P. Mitra, Multiple wave-vector extensions of the {NMR} pulsed-field-gradient
  spin-echo diffusion measurement, Phys Rev B Condens Matter 51~(21) (1995)
  15074--15078.

\bibitem{Shemesh2016}
N.~Shemesh, S.~N. Jespersen, D.~C. Alexander, Y.~Cohen, I.~Drobnjak, T.~B.
  Dyrby, J.~Finsterbusch, M.~A. Koch, T.~Kuder, F.~Laun, M.~Lawrenz,
  H.~Lundell, P.~P. Mitra, M.~Nilsson, E.~{\"O}zarslan, D.~Topgaard, C.-F.
  Westin, Conventions and nomenclature for double diffusion encoding {NMR} and
  {MRI}, Magn Reson Med 75~(1) (2016) 82--7.
\newblock \href {http://dx.doi.org/10.1002/mrm.25901}
  {\path{doi:10.1002/mrm.25901}}.

\bibitem{Eriksson2013}
S.~Eriksson, S.~Lasic, D.~Topgaard, Isotropic diffusion weighting in {PGSE NMR}
  by magic-angle spinning of the q-vector, Journal of Magnetic Resonance 226
  (2013) 13 -- 18.
\newblock \href {http://dx.doi.org/https://doi.org/10.1016/j.jmr.2012.10.015}
  {\path{doi:https://doi.org/10.1016/j.jmr.2012.10.015}}.

\bibitem{Lasic2014}
S.~Lasi\v{c}, F.~Szczepankiewicz, S.~Eriksson, M.~Nilsson, D.~Topgaard,
  Microanisotropy imaging: quantification of microscopic diffusion anisotropy
  and orientational order parameter by diffusion {MRI} with magic-angle
  spinning of the q-vector, Frontiers in Physics 2 (2014) 11.
\newblock \href {http://dx.doi.org/10.3389/fphy.2014.00011}
  {\path{doi:10.3389/fphy.2014.00011}}.

\bibitem{Szczepankiewicz2015}
F.~Szczepankiewicz, S.~Lasi\v{c}, D.~van Westen, P.~C. Sundgren, E.~Englund,
  C.-F. Westin, F.~St{\aa}hlberg, J.~L{\"a}tt, D.~Topgaard, M.~Nilsson,
  Quantification of microscopic diffusion anisotropy disentangles effects of
  orientation dispersion from microstructure: applications in healthy
  volunteers and in brain tumors, Neuroimage 104 (2015) 241--52.
\newblock \href {http://dx.doi.org/10.1016/j.neuroimage.2014.09.057}
  {\path{doi:10.1016/j.neuroimage.2014.09.057}}.

\bibitem{Jespersen2012_dPFG}
S.~N. Jespersen, Equivalence of double and single wave vector diffusion
  contrast at low diffusion weighting, NMR Biomed 25~(6) (2012) 813--8.
\newblock \href {http://dx.doi.org/10.1002/nbm.1808}
  {\path{doi:10.1002/nbm.1808}}.

\bibitem{Komlosh2007}
M.~E. Komlosh, F.~Horkay, R.~Z. Freidlin, U.~Nevo, Y.~Assaf, P.~J. Basser,
  Detection of microscopic anisotropy in gray matter and in a novel tissue
  phantom using double {Pulsed Gradient Spin Echo MR}, J Magn Reson 189~(1)
  (2007) 38--45.
\newblock \href {http://dx.doi.org/10.1016/j.jmr.2007.07.003}
  {\path{doi:10.1016/j.jmr.2007.07.003}}.

\bibitem{Shemesh2012}
N.~Shemesh, D.~Barazany, O.~Sadan, L.~Bar, Y.~Zur, Y.~Barhum, N.~Sochen,
  D.~Offen, Y.~Assaf, Y.~Cohen, Mapping apparent eccentricity and residual
  ensemble anisotropy in the gray matter using angular
  double-pulsed-field-gradient {MRI}, Magn Reson Med 68~(3) (2012) 794--806.
\newblock \href {http://dx.doi.org/10.1002/mrm.23300}
  {\path{doi:10.1002/mrm.23300}}.

\bibitem{Lawrenz2013}
M.~Lawrenz, J.~Finsterbusch, Double-wave-vector diffusion-weighted imaging
  reveals microscopic diffusion anisotropy in the living human brain, Magn
  Reson Med 69~(4) (2013) 1072--1082.
\newblock \href {http://dx.doi.org/10.1002/mrm.24347}
  {\path{doi:10.1002/mrm.24347}}.

\bibitem{Lawrenz2019}
M.~Lawrenz, J.~Finsterbusch, Detection of microscopic diffusion anisotropy in
  human cortical gray matter in vivo with double diffusion encoding, Magn Reson
  Med 81~(2) (2019) 1296--1306.
\newblock \href {http://dx.doi.org/10.1002/mrm.27451}
  {\path{doi:10.1002/mrm.27451}}.

\bibitem{Jespersen2013}
S.~N. Jespersen, H.~Lundell, C.~K. Sonderby, T.~B. Dyrby, Orientationally
  invariant metrics of apparent compartment eccentricity from double pulsed
  field gradient diffusion experiments, {NMR IN BIOMEDICINE} {26}~({12})
  ({2013}) {1647--1662}.
\newblock \href {http://dx.doi.org/{10.1002/nbm.2999}}
  {\path{doi:{10.1002/nbm.2999}}}.

\bibitem{Lampinen2017}
B.~Lampinen, F.~Szczepankiewicz, J.~M{\aa}rtensson, D.~van Westen, P.~C.
  Sundgren, M.~Nilsson, Neurite density imaging versus imaging of microscopic
  anisotropy in diffusion {MRI: A} model comparison using spherical tensor
  encoding, Neuroimage 147 (2017) 517--531.
\newblock \href {http://dx.doi.org/10.1016/j.neuroimage.2016.11.053}
  {\path{doi:10.1016/j.neuroimage.2016.11.053}}.

\bibitem{Jespersen2014}
S.~N. Jespersen, H.~Lundell, C.~K. S{\o}nderby, T.~B. Dyrby, Commentary on
  ''microanisotropy imaging: quantification of microscopic diffusion anisotropy
  and orientation of order parameter by diffusion mri with magic-angle spinning
  of the q-vector'', Frontiers in Physics 2~(28).
\newblock \href {http://dx.doi.org/10.3389/fphy.2014.00028}
  {\path{doi:10.3389/fphy.2014.00028}}.

\bibitem{Jespersen2019}
S.~N. Jespersen, J.~L. Olesen, A.~Ianu\c{s}, N.~Shemesh, Effects of nongaussian
  diffusion on ``isotropic diffusion'' measurements: {An} ex-vivo microimaging
  and simulation study, J Magn Reson 300 (2019) 84--94.
\newblock \href {http://dx.doi.org/10.1016/j.jmr.2019.01.007}
  {\path{doi:10.1016/j.jmr.2019.01.007}}.

\bibitem{Dhital2018_iso}
B.~Dhital, E.~Kellner, V.~G. Kiselev, M.~Reisert, The absence of restricted
  water pool in brain white matter, Neuroimage 182 (2018) 398--406.
\newblock \href {http://dx.doi.org/10.1016/j.neuroimage.2017.10.051}
  {\path{doi:10.1016/j.neuroimage.2017.10.051}}.

\bibitem{Tax2020_dot}
C.~M. Tax, F.~Szczepankiewicz, M.~Nilsson, D.~K. Jones, The dot-compartment
  revealed? {D}iffusion {MRI} with ultra-strong gradients and spherical tensor
  encoding in the living human brain, Neuroimage 210 (2020) 116534.
\newblock \href {http://dx.doi.org/10.1016/j.neuroimage.2020.116534}
  {\path{doi:10.1016/j.neuroimage.2020.116534}}.

\bibitem{Aslund2009}
I.~Aslund, A.~Nowacka, M.~Nilsson, D.~Topgaard, Filter-exchange {PGSE NMR}
  determination of cell membrane permeability, J Magn Reson 200~(2) (2009)
  291--295.
\newblock \href {http://dx.doi.org/10.1016/j.jmr.2009.07.015}
  {\path{doi:10.1016/j.jmr.2009.07.015}}.

\bibitem{Lasic2011_exchange}
S.~Lasi{\v c}, M.~Nilsson, J.~L{\"a}tt, F.~St{\aa}hlberg, D.~Topgaard, Apparent
  exchange rate mapping with diffusion mri, Magnetic Resonance in Medicine
  66~(2) (2011) 356--365.
\newblock \href
  {http://arxiv.org/abs/https://onlinelibrary.wiley.com/doi/pdf/10.1002/mrm.22782}
  {\path{arXiv:https://onlinelibrary.wiley.com/doi/pdf/10.1002/mrm.22782}},
  \href {http://dx.doi.org/10.1002/mrm.22782} {\path{doi:10.1002/mrm.22782}}.

\bibitem{Lampinen2016}
B.~Lampinen, F.~Szczepankiewicz, D.~van Westen, E.~Englund, P.~C~Sundgren,
  J.~L{\"a}tt, F.~St{\aa}hlberg, M.~Nilsson, Optimal experimental design for
  filter exchange imaging: {A}pparent exchange rate measurements in the healthy
  brain and in intracranial tumors, Magn Reson Med\href
  {http://dx.doi.org/10.1002/mrm.26195} {\path{doi:10.1002/mrm.26195}}.

\bibitem{Lasic2016}
S.~Lasi{\v{c}}, S.~Oredsson, S.~C. Partridge, L.~H. Saal, D.~Topgaard,
  M.~Nilsson, K.~Bryskhe, Apparent exchange rate for breast cancer
  characterization, NMR Biomed 29~(5) (2016) 631--9.
\newblock \href {http://dx.doi.org/10.1002/nbm.3504}
  {\path{doi:10.1002/nbm.3504}}.

\bibitem{Dhital2019}
B.~Dhital, M.~Reisert, E.~Kellner, V.~G. Kiselev, Intra-axonal diffusivity in
  brain white matter, NeuroImage 189 (2019) 543--550.
\newblock \href {http://dx.doi.org/10.1016/j.neuroimage.2019.01.015}
  {\path{doi:10.1016/j.neuroimage.2019.01.015}}.

\bibitem{Jensen2016}
J.~H. Jensen, G.~Russell~Glenn, J.~A. Helpern, Fiber ball imaging, Neuroimage
  124~(Pt A) (2016) 824--833.
\newblock \href {http://dx.doi.org/10.1016/j.neuroimage.2015.09.049}
  {\path{doi:10.1016/j.neuroimage.2015.09.049}}.

\bibitem{Huerlimann98}
M.~D. H{\"{u}}rlimann, {Effective Gradients in Porous Media Due to
  Susceptibility Differences}, J Magn Reson 131 (1998) 232--240.
\newblock \href {http://dx.doi.org/10.1006/jmre.1998.1364}
  {\path{doi:10.1006/jmre.1998.1364}}.

\bibitem{Kiselev2019}
V.~G. Kiselev, {L}armor frequency in heterogeneous media, J Magn Reson 299
  (2019) 168--175.
\newblock \href {http://dx.doi.org/10.1016/j.jmr.2018.12.008}
  {\path{doi:10.1016/j.jmr.2018.12.008}}.

\bibitem{Kiselev2002}
V.~G. Kiselev, D.~S. Novikov, Transverse {NMR} relaxation as a probe of
  mesoscopic structure, Physical Review Letters 89 (2002) 278101.

\bibitem{Novikov2008}
D.~S. Novikov, V.~G. Kiselev, Transverse {NMR} relaxation in magnetically
  heterogeneous media, J Magn Reson 195~(1) (2008) 33--39.

\bibitem{Does99}
M.~D. Does, J.~Zhong, J.~C. Gore, In vivo measurement of {ADC} change due to
  intravascular susceptibility variation, Magnetic Resonance in Medicine 41~(2)
  (1999) 236--240.
\newblock \href
  {http://dx.doi.org/10.1002/(SICI)1522-2594(199902)41:2<236::AID-MRM4>3.0.CO;2-3}
  {\path{doi:10.1002/(SICI)1522-2594(199902)41:2<236::AID-MRM4>3.0.CO;2-3}}.

\bibitem{Zhong91}
J.~Zhong, R.~P. Kennan, J.~C. Gore, Effects of susceptibility variations on
  {NMR} measurements of diffusion, Journal of Magnetic Resonance (1969) 95~(2)
  (1991) 267 -- 280.
\newblock \href
  {http://dx.doi.org/http://dx.doi.org/10.1016/0022-2364(91)90217-H}
  {\path{doi:http://dx.doi.org/10.1016/0022-2364(91)90217-H}}.

\bibitem{Kiselev2004}
V.~G. Kiselev, Effect of magnetic field gradients induced by microvasculature
  on {NMR} measurements of molecular self-diffusion in biological tissues, J
  Magn Reson 170 (2004) 228 -- 235.

\bibitem{Novikov2018_ackerman}
D.~S. Novikov, M.~Reisert, V.~G. Kiselev, Effects of mesoscopic susceptibility
  and transverse relaxation on diffusion nmr, Journal of Magnetic Resonance 293
  (2018) 134 -- 144.
\newblock \href {http://dx.doi.org/https://doi.org/10.1016/j.jmr.2018.06.007}
  {\path{doi:https://doi.org/10.1016/j.jmr.2018.06.007}}.

\end{thebibliography}

\end{document}